\documentclass[reprint,amsmath,amssymb,aps,showkeys]{revtex4-2}

\usepackage{slashed}

\usepackage{graphicx}
\usepackage{dcolumn}
\usepackage{bm}
\usepackage{hyperref}
\usepackage[mathlines]{lineno}
\usepackage{xcolor}
\usepackage[makeroom]{cancel}

\newcommand\RR{\mathbb{R}}
\newcommand\CC{\mathbb{C}}

\renewcommand\H{\mathcal{H}}

\newcommand\U{\mathcal{U}}
\newcommand\SO{\mathcal{SO}}
\newcommand\SL{\mathcal{SL}}

\newcommand\K{\mathcal{K}}

\newcommand\GL{\mathcal{GL}}
\newcommand\D{\mathcal{D}}

\newcommand\rarrow{\rightarrow}

\newcommand\LieH{\mathfrak{h}}

\newcommand\LieK{\mathfrak{K}}

\renewcommand\t{\tilde}

\renewcommand\b{\bar }

\newcommand\bs{\boldsymbol}

\renewcommand\-{^{-1}}

\newcommand\ad{\text{ad}}

\DeclareMathOperator{\Diff}{Diff}

\begin{document}


\title{Unconventional Supersymmetry via the Dressing Field Method}

\author{J. François}
\email{jordan.francois@uni-graz.at}
\affiliation{University of Graz (Uni Graz), 
Heinrichstraße 26/5, 8010 Graz, Austria, and \\
Masaryk University (MUNI), 
Kotlářská 267/2, Veveří, Brno, Czech Republic, and \\
Mons University (UMONS), 
20 Place du Parc, 7000 Mons, Belgium. 
}

\author{L. Ravera}
\email{lucrezia.ravera@polito.it}
\affiliation{Politecnico di Torino (PoliTo),
C.so Duca degli Abruzzi 24, 10129 Torino, Italy, and \\
Istituto Nazionale di Fisica Nucleare (INFN), Section of Torino,
Via P. Giuria 1, 10125 Torino, Italy, and \\
Grupo de Investigación en Física Teórica (GIFT),
Universidad Cat\'{o}lica De La Sant\'{i}sima Concepci\'{o}n, Concepción, Chile.
}

\date{\today}

\begin{abstract}
We re-construe \emph{unconventional supersymmetry}, a notion introduced by Alvarez-Valenzuela-Zanelli (AVZ), as an attempt to  use the framework of supersymmetric field theory to describe fermionic matter fields and bosonic gauge fields in a unified way, as parts of a single superconnection. 
It~hinges upon the so-called \emph{matter ansatz}. 
Unfortunately, the formal and conceptual status of the ansatz has remained unclear, preventing unconventional supersymmetry to be used in a principled way as a general approach beyond the model in which it was first considered.
In this letter, we lift this restriction by showing that the ansatz is a special case of the \emph{Dressing Field Method}, a new systematic tool to exhibit the gauge-invariant content of general-relativistic gauge field theories. 

\end{abstract}

\keywords{Unconventional supersymmetry, Dressing Field Method, Gauge invariance, Relationality.}

\maketitle


\section{Introduction}\label{Introduction}

Supersymmetric field theory, as conventionally applied within high energy physics, implies a doubling of the number of fundamental fields and particles whereby each known particle has a superpartner of opposite statistics: fermionic (matter) fields have bosonic partners, bosonic (gauge and Higgs) fields have fermionic partners. 
This view can only be accommodated with empirical data via
the notion of (spontaneous) supersymmetry (\emph{susy}) breaking.
Still, until now, supersymmetric particles did not show up in colliders.

This nonetheless does not spell doom for the framework of supersymmetric field theory, which by itself does not require a matching of bosonic and fermionic degrees of freedom (d.o.f.) \cite{Sohnius:1985qm}, and has been shown to provide original new insights into established physics
such as
QCD \cite{Armoni:2010zz} and condensed matter physics \cite{sourlas1985,Efetov:1997fw,Ma:2021dua}. 
Furthermore, one may recall that Berezin pioneered the introduction of supergeometry in physics to describe fermions and their peculiar statistics \cite{Berezin-Marinov1977}.
A program in dormancy that still awaits to yield its full potential. 

In that spirit, ``unconventional supersymmetry", \emph{ususy} hereafter,  introduced by Alvarez-Valenzuela-Zanelli (AVZ) 
\cite{Alvarez:2011gd,Alvarez:2013tga},
is best understood as a way to use the framework of supersymmetric field theory to describe fermionic matter fields and bosonic gauge fields in a unified way: both being parts of one and the same superconnection \cite{Guevara:2016rbl,Zanelli:2017hty,Alvarez:2021zhh,Alvarez:2023auf} (see also 
\cite{Andrianopoli:2018ymh, Andrianopoli:2020zbl, Alvarez:2022wcj}). 
This is achieved through what was called the ``matter ansatz" in \cite{Alvarez:2021zhh}. 
An interesting parallel is to be drawn  with proposals to use the framework of non-commutative geometry -- either derivation-based, or via Lie algebroids, or yet via spectral triples -- to describe gauge fields and the Higgs field as unified in a single non-commutative connection \cite{Dubois-Violette_kerner_Madore1990a,Dubois-Violette-Kerner-Madore1990b,Connes-Lott1991, Masson-Lazz2013,Masson-Lazz}.
A commonality that is not so surprising as supergeometry, i.e. $\mathbb Z_2$-graded geometry, is the ``mildest" form of non-commutative geometry. 

Yet, the technical and conceptual status of the ansatz remained unclear, preventing ususy to be used in a principled way as a general approach outgrowing 
the so-called AVZ model
in which it was first considered.
We here lift this restriction, by showing that the ususy matter ansatz is subsumed by 
the  \emph{Dressing Field Method} (DFM). 

The DFM, introduced in \cite{GaugeInvCompFields}, is a systematic approach to construct gauge-invariant variables in Gauge Field Theory (GFT).
It is best understood within the formalism of the bundle differential geometry of field space \cite{Francois2021,FrancoisAndre:2023jmj,Zajac2023}, but it also has a more broadly accessible field-theoretic framing \cite{Berghofer-et-al2023}. 
Unfortunately, in the latter context dressings are often conflated with mere  gauge-fixings,   generating misconceptions regarding the physics.
This usually happens when the basic 
definition of a gauge group is overlooked, regrettably not a rare occurrence in the GFT literature, as detailed in \cite{Berghofer:2024plf,Francois:2024laf}.
\mbox{Integral} to the DFM is its  natural \emph{relational} interpretation:  gauge-invariance being achieved by realising the \emph{physical} d.o.f. representing \emph{relations} among \emph{bare} (gauge-variant) d.o.f.  \cite{FrancoisAndre:2023jmj,Francois:2024rdm}.
Relationality, thus understood, is
 the core paradigmatic conceptual insight of the 
general-relativistic gauge field theory framework \cite{Francois:2024vlr}, which includes
supersymmetric field theories.
It is thus surprising that the relational character of the latter has been so completely overlooked in the literature, until it was first explicitly signaled in \cite{Francois:2024laf}, where the DFM is shown necessary to build the Rarita-Schwinger and gravitino fields. 

\noindent
In this paper, we thus pursue two complementary goals:
\begin{enumerate}
    \item To elucidate the technical and conceptual status of the ususy matter ansatz as a case of dressing, thereby allowing ususy to be used in a principled way as an approach to a unified description of gauge and matter fields. 
    \item To bring the attention of a broader community to the DFM, as a new systematic approach to reformulate general-relativistic gauge field theories in a manifestly gauge-invariant way, thereby automatically extracting their physical content, and highlighting its \emph{relational} character.
\end{enumerate}
A prerequisite to achieve both objectives is of course that we introduce the basics of the DFM. 
Achieving 1 requires us to be concrete: we thus opt to illustrate our claim via the first and simplest case, the AVZ ususy model. 
We endeavour to present it in a way that gives a clean template to showcase a clear application of the DFM, so that the reader can easily export the approach to any gauge/supersymmetric field theoretic model of their choosing (from high energy to condensed matter physics), achieving 2.
It will also make clear how to obtain other viable ususy models via DFM, decisively securing 1. 

The remaining of the letter is thus organized as follows.
In Section \ref{The AVZ model of ususy} we provide a streamlined presentation of the AVZ ususy model, relying on a compact matrix notation, and insisting on the proper logic: kinematics and symmetries first, then only the dynamics. 
We highlight a key point regarding the definition of a gauge group. 
Almost always overlooked, it is key to understand the DFM -- and GFT more broadly, we  argue. 
In Section \ref{Matter ansatz via the DFM}, we explain the DFM and show that the matter ansatz is the natural end result of a dressing operation, highlighting the relational character of the invariant variables thus produced. 
In Section \ref{Residual gauge symmetries} we show how the residual gauge symmetries of the model are easily handled via the DFM framework, also  clarifying the status of the so-called Nieh-Yan-Weyl  symmetry \cite{Nieh:1981xk,Chandia:1997hu,Kreimer:1999yp,Hughes:2012vg,Parrikar:2014usa}. 
Section \ref{Conclusions} gathers  our results and we
discuss future developments. 
In Appendix \ref{Ususy via DFM in any dimension} we show how to build ususy models via DFM in  dimension $d>3$.

\section{The AVZ model of ususy}\label{The AVZ model of ususy}

The model, introduced in 
\cite{Alvarez:2011gd},
is a Chern-Simons (CS) theory in $2 + 1$ dimensions for the supergroup $\rm{OSp}(2|2)$. 
In the following, we shall give a compact and synthetic description of its kinematics and  dynamics.

\subsection{Kinematics of the model}
\label{Kinematic}

The field-theoretic presentation of the AVZ model  is based on the differential geometry of a principal superbundle  
with structure supergroup $H=\rm{OSp}(2|2)$, with associated Lie superalgebra  $\LieH:=\mathfrak{osp}(2|2)$. 
Already we must stress a basic but  important point regarding the definition of 
the gauge group of a theory: it derives from the  group of vertical automorphisms of the principal  bundle underlying the kinematics of the theory. 
In the case at hand, 
the vertical automorphism group of the underlying  superbundle induces, at the field-theoretic level -- i.e. on the base (bosonic) spacetime manifold $M$ -- the gauge group 
$\mathcal H=\mathcal{OS}p(2|2) := \{ g, g': U \subset M \rarrow H\,|\, {g}^{\prime\, g} = \text{Conj}(g\-)\, g':=g\- g' g \}$, i.e. \emph{a gauge group acts on its own elements via the conjugation action}. Correspondingly, the gauge algebra is
Lie$\mathcal H=$Lie$\mathcal{OS}p(2|2) := \{ \lambda, \lambda': U  \rarrow \LieH\,|\, \delta_\lambda \lambda' = \ad(-\lambda)\, \lambda'=[\lambda', \lambda] \}$, i.e. the gauge algebra elements are $\ad$-tensorial 0-forms.
We have  $g(x)=e^{\lambda(x)}$. 
We shall use the compact matrix notation
\begin{align}
\label{gaugeparam}
    \lambda = \begin{pmatrix}
        \beta  & 
\ \tfrac{1}{\sqrt{\ell}} \varepsilon \\
      - \tfrac{1}{\sqrt{\ell}} \bar{\varepsilon}   & \   \alpha \otimes J
    \end{pmatrix} ,
\end{align}
where $\beta$ takes values in $\mathfrak{sp}(2, \RR)\simeq\mathfrak{sl}(2, \RR)\simeq \mathfrak{spin}(1, 2)$, $\varepsilon \in \CC^2$ is a spinor field with $\b \varepsilon := \varepsilon^\dagger \gamma_0 =- \varepsilon^\dagger i J$, 
$\alpha$ is $\mathfrak{u}(1)=i\mathbb{R}$-valued and $\alpha \otimes J$ is to be treated as a scalar in the matrix $\lambda$ even though $J$ is the symplectic matrix (in component $J\sim \epsilon_{AB}$), acting on $\varepsilon$ as $J\varepsilon$ and on $\bar \varepsilon$ as $\b \varepsilon J$. 
The \emph{defining} transformation property of the infinitesimal gauge parameters is then
\begin{align}
\label{gaugeparam-trsf}
& \delta_\lambda\lambda' 
= \begin{pmatrix}
    \delta_\lambda \beta'  & 
 \ \tfrac{1}{\sqrt{\ell}} \delta_\lambda \varepsilon' \\[2mm]
- \tfrac{1}{\sqrt{\ell}} \delta_\lambda   \bar{\varepsilon}'   &\    \delta_\lambda \alpha' \otimes J
    \end{pmatrix} \\[1mm]
&{\scriptsize{:= \begin{pmatrix}
 [\beta',\beta] -\tfrac{1}{\ell} \left(\varepsilon' \bar\varepsilon - \varepsilon \bar\varepsilon' \right) &  \tfrac{1}{\sqrt{\ell}}\left[(\beta'-\alpha'\otimes J)\varepsilon - (\beta-\alpha\otimes J) \varepsilon' \right]\\[2mm]
 *
  & \tfrac{1}{2\ell} \mathrm{Tr} (\varepsilon' \bar \varepsilon - \varepsilon \bar \varepsilon')\otimes J
\end{pmatrix} ,}}\notag
\end{align}
the factor $\tfrac{1}{2}$ in the (2,2) entry being given by prescription. 
This then encodes  the commutators between generators of $\LieH=\mathfrak{osp}(2|2)$, in terms of which a Lie$\H$ element reads
$\lambda(x)=\tfrac{1}{2}\beta^{ab}(x) \mathbb{J}_{ab}+\bar{\mathbb{Q}}\varepsilon(x)-\bar{\varepsilon}(x) \mathbb{Q}+\alpha(x) \mathbb{T}$,
with $\mathbb{J}$, $\mathbb{Q}$, $\bar{\mathbb{Q}}$, and $\mathbb{T}$ the generators of  Lorentz, susy and $\mathrm{U}(1)$ transformations, respectively.

From a basic gauge theoretic perspective, a generic gauge field should be an $\LieH$-valued connection 1-form $\mathbb{A}_\mathfrak{osp}$. 
What makes the AVZ model ``unconventional", from a supersymmetric standpoint, is that the spinor-valued 1-form $\psi$, gauge potential of supersymmetry
that should feature in $\mathbb{A}_\mathfrak{osp}$ as $\b{\mathbb{Q}} \psi - \b\psi \mathbb{Q}$, is replaced by a composite
\begin{align}
\label{matterans}
    \psi = i \gamma_a e^a \chi ,
\end{align}
 where ${e^a}= {e^a}_\mu dx^{\mu}$ is the soldering form on $M$ inducing a metric with signature  $(+,-,-)$,
$\gamma^a$
are the gamma matrices, and $\chi$  is a spin-$1/2$ (Dirac) field.
The substitution \eqref{matterans} is called  the ``\emph{matter ansatz}"  \cite{Alvarez:2021zhh}.
The superconnection adopted in the AVZ model is  thus the 1-form $\mathbb{A}=\mathbb{A}_\mu dx^\mu
=\frac{1}{2}\omega^{ab} \mathbb{J}_{ab}+ i\, \bar{\mathbb{Q}} \,\gamma\chi+i\,\bar \chi \gamma \,\mathbb{Q}+A\, \mathbb{T}$.
In the above compact matrix form,
\begin{align}
\label{connection}
    \mathbb{A} 
= \begin{pmatrix}
  \omega &\  \tfrac{i}{\sqrt{\ell}}\,\gamma 
 \chi \\[2mm]
   \tfrac{i}{\sqrt{\ell}}\,\bar \chi \gamma   & \  A \otimes J
\end{pmatrix} ,
\end{align}
where 
$\gamma$ denotes the gamma matrices 1-form $\gamma =\gamma_\mu dx^{\mu}=\gamma_a {e^a}_\mu dx^{\mu}=\gamma_a e^a$. 
Our convention for the gamma matrices in $2+1$ dimensions is $\gamma_0=\sigma_2=-i J$, $\gamma_1=-i\sigma_1$, $\gamma_2=-i \sigma_3$. The parameter $1/\ell$ is related to the negative cosmological constant $\Lambda$ by $\Lambda=-1/\ell^2$.  
The associated curvature 2-form is, by Cartan structure equation,
$\mathbb F= d\mathbb{A} + \tfrac{1}{2}[\mathbb{A}, \mathbb{A}] =d\mathbb{A} + \mathbb{A}^2$. 
In~matrix form,
\begin{equation}
\begin{aligned}
\label{curvature}
    \mathbb{F} 
=&\, \begin{pmatrix}
    \Omega  & \ \tfrac{i}{\sqrt{\ell}} \varrho \\[2mm]
    *    & \,  \t F \otimes J
  \end{pmatrix} \\[2mm]
:=&\, \begin{pmatrix}
    R -\tfrac{1}{\ell}  \gamma \chi  \,  \b\chi \gamma  &\  \tfrac{i}{\sqrt{\ell}} \left(\gamma_a T^a \chi - \gamma \nabla \chi \right) \\[2mm]
    *    & \, \left( dA - \tfrac{1}{2\ell}\b\chi\gamma \, \gamma \chi \right) \otimes J 
  \end{pmatrix} ,
\end{aligned}
\end{equation}
where $R:=d\omega + \omega^2$ and $T^a := de^a + {\omega^a}_b e^b = {T^a}_{bc} e^b \wedge e^c$ are  the spacetime curvature and torsion 2-forms, respectively, and, in this notation, $\nabla \chi := d\chi + \omega \chi - A\otimes J \chi$.
The curvature $\mathbb F$ satisfies the Bianchi identity $D^{\mathbb A}\mathbb{F}= d\mathbb{F}+[\mathbb{A}, \mathbb{F}]\equiv 0$, where   $D^{\mathbb A}$ denotes the covariant derivative with respect to the full $\mathbb A$, here adapted to  $\ad$-tensorial forms. 

The $\mathcal{OS}p(2|2)$-transformations of the connection 
is $\mathbb{A}^g=g^{-1}\mathbb{A} g + g^{-1}dg$, which infinitesimally restricts as
 \mbox{$\delta_\lambda \mathbb{A} = D^{\mathbb A} \lambda := d \lambda + \left[\mathbb{A},\lambda \right]$}. 
In matrix form,
\begin{align}
\label{infgaugetr}
&\delta_\lambda \mathbb{A} 
= \begin{pmatrix}
    \delta_\lambda \omega  & 
 \ \tfrac{i}{\sqrt \ell} \delta_\lambda (\gamma\chi)  \\[2mm]
 \tfrac{i}{\sqrt \ell} \delta_\lambda   (\b{\chi}\gamma )  &\    \delta_\lambda A \otimes J
    \end{pmatrix} 
\\[2mm]
&={\scriptsize{\begin{pmatrix}
 d\beta + [\omega,\beta] - \tfrac{i}{\ell}(\gamma\chi \,\bar\varepsilon + \varepsilon \,\b\chi\gamma) 
 & \tfrac{1}{\sqrt \ell}\left[
 \nabla \varepsilon
 - (\beta-\alpha \otimes J) i \gamma\chi \, \right] \\[2mm]
 *
  & \left( d\alpha+ \tfrac{i}{2\ell} \mathrm{Tr} (\gamma\chi\, \bar \varepsilon + \varepsilon \, \b\chi\gamma) \right) \otimes J 
\end{pmatrix} .}} \notag 
\end{align}
The curvature transforms tensorially: $\mathbb{F}^g=g\- \mathbb{F}g$, so 
$\delta_\lambda \mathbb{F}= [\mathbb{F}, \lambda]$.

\subsection{Dynamics of the  model}
\label{Dynamics}

The Lagrangian of the theory is a CS 3-form, which is the simplest Lagrangian one can think of for the connection $\mathbb{A}$ in $2+1$ dimensions. It reads
\begin{align}\label{CSLagr}
    L (\mathbb A)= \mathrm{sTr} \left(\mathbb{A} d \mathbb{A} + \frac{2}{3} \mathbb{A}^3 \right) 
      = \mathrm{sTr} \left(\mathbb{A} \mathbb{F} - \frac{1}{3} \mathbb{A}^3 \right)
    .
\end{align}
It is \emph{not} $\mathcal{OS}p(2|2)$-invariant, \emph{nor} quasi-invariant (CS forms never are under gauge group transformations, see \cite{Francois2021}), and is merely quasi-invariant (transforms via a $d$-exact term) under Lie$\mathcal{OS}p(2|2)$. 
The latter still suffices to ensure that the field equations are $\mathcal{OS}p(2|2)$-covariant: indeed they are, since -- as is well-known in  CS theories -- given by the flatness condition $\mathbb{ F}=0$, which unfolds in an obvious way as field equations for $\omega$, $A$ and $\chi$ by \eqref{curvature}. 
In components they read, explicitly:
\begin{equation}
\label{unfolded-field-eq}
\begin{aligned}
    \Omega^{ab} &:= R^{ab} + \frac{2}{\ell} \bar\chi_A \chi_A e^a \wedge e^b =0 , \\
    \tilde{F} &:= dA -\frac{i}{2\ell} \bar{\chi}_A \gamma^c \chi_B \epsilon_{AB} \epsilon_{abc} e^a \wedge e^b = 0 ,  \\
    \phantom{\Rightarrow} & \quad \gamma_a T^a \chi_A - \gamma \nabla \chi_A =0  \\
    \Rightarrow & \quad \gamma_a {T^a}_{\mu \nu} \chi_A - \gamma_{[\mu} \nabla_{\nu]} \chi_A = 0 . 
\end{aligned}
\end{equation}
Contracting the last equation with $\gamma^{\mu \nu}:=\tfrac{1}{2}\left[\gamma^\mu,\gamma^\nu\right]$, we get the massive Dirac equation 
\begin{align}
\label{diraceq}
    \slashed{\nabla} \chi_A = \left( {T^a}_a - 3 i \tau \right) \chi_A ,
\end{align}
where we used $\gamma^{abc}=i\epsilon^{abc}$ and defined ${T^a}_a:=\gamma^b {T^a}_{ab}$ and $\tau:= \tfrac{1}{3!}\epsilon^{abc}T_{abc}$. 
One could further simplify \eqref{diraceq} by expressing $\omega$ in terms of the torsion.
\medskip

This model was shown to have important applications in condensed matter physics, describing in particular graphene-like systems near the Dirac points in a
generic spatial lattice with curvature and torsion \cite{Iorio:2011yz,Iorio:2013ifa,Iorio:2014pwa,Iorio:2014nda,Iorio:2018agc,Ciappina:2019qgj,Andrianopoli:2019sip,Iorio:2020tuc,Acquaviva:2022yiq,Iorio:2022ave,Iorio:2022jrg}.
A proof of concept that gauge and matter fields can be successfully described in a unified way within supersymmetric field theory. 
Only, the meaning of the  ansatz \eqref{matterans}
remained unclear, limiting the potential reach of the result.
In \cite{Valenzuela:2022gbk,Valenzuela:2023aoa},  \eqref{matterans} was said to correspond to a ``projection" of the vector-spinor $\psi$, in which its gamma-traceless component vanishes, while later, in \cite{Andrianopoli:2019sqe}, it was claimed to be the result of a gauge fixing -- more precisely, a BRST-covariant gauge fixing of the odd symmetries (see also \cite{Andrianopoli:2021sdx}).
We settle the issue by showing in the next section that \eqref{matterans} actually results from a dressing operation.

\section{Matter ansatz via the DFM}\label{Matter ansatz via the DFM}

We first briefly describe the Dressing Field Method, and then  show how it yields the ususy  matter ansatz. 

\subsection{The DFM and its infinitesimal version}
\label{The DFM and its infinitesimal version}

The DFM \cite{GaugeInvCompFields,Francois2021,FrancoisAndre:2023jmj,Zajac2023,Francois:2024laf,FrancoisAndre:2023jmj,Francois:2024vlr,Francois:2024rdm,Francois:2024laf} is a systematic tool to produce gauge-invariants out of the field space $\Phi=\{\phi\}$ of a gauge theory with gauge group $\H$ whose action on $\Phi$ defines gauge transformations:  $\phi \mapsto \phi^{g}$. 

Suppose $\K \subseteq \H$ is a gauge subgroup, corresponding to the rigid subgroup $K\subseteq H$. The DFM relies on identifying a \emph{$\Phi$-dependent $\K$-dressing field}, i.e. a map
\begin{equation}
\label{Field-dep-dressing}
\begin{aligned}
u :  \Phi \ &\rarrow\  \D r[K, \K], \\
    \phi \  &\mapsto\  u=u[\phi], \\
    \phi^k &\mapsto\ u^k:=u[\phi^k]=k\-u[\phi], \ \ \forall k\in \K,
\end{aligned} 
\end{equation}
where $\D r[K, \K]:=\{u: U\subset M \rarrow K\,|\, u^k=k\-u\}$ is the space of ($\Phi$-independent) dressing fields.
From it, one can systematically build the $\K$-\emph{invariant dressed fields}
by the surjective map
\begin{equation}
\label{Dressed-field}
\begin{aligned}
 \Phi \ &\rarrow\  \Phi^u, \\
    \phi \  &\mapsto\  \phi^u=\phi^{u[\phi]}, \\
    \phi^k &\mapsto\ (\phi^k)^{u^k}:=(\phi^k)^{k\-u[\phi]}=\phi^{u[\phi]}.
\end{aligned} 
\end{equation} 
For $\phi=A$ a gauge potential $\H$-transforming as $A^g:=g\- A g+g\-dg$, its  $\K$-dressing is $A^u:=u\-Au +u\-du$. It~is easily seen to be $\K$-invariant: $(A^u)^k :=(A^k)^{u^k}=A^u$.

If one is interested, as we shall be here, in invariance at first order, i.e. under infinitesimal gauge transformations Lie$\H$, one may  linearise the above:  Defining a Lie$\K$-dressing field as
\begin{equation}
\label{dressingfieldtr}
\begin{aligned}
&\upsilon=\upsilon[\phi]: U\subset M \rarrow \LieK=\text{Lie}K, \\[1mm]
& \text{s.t. } \quad \delta_\lambda \upsilon :=\upsilon[\delta_\lambda \phi] \approx -\lambda , \ \ \forall\lambda \in \text{Lie}\K,
\end{aligned}
\end{equation}
where in the defining transformation law one is to neglect higher-order terms polynomial in $\lambda$ and $\upsilon$.
Then, one defines the \emph{perturbatively dressed fields} 
\begin{align}
\label{pert-dressed-fields}
\phi^\upsilon:= \phi + \b\delta_\upsilon \phi,
\end{align}
where $\b\delta_\upsilon \phi$ mimics the functional expression of the Lie$\H$ gauge  transformation $\delta_\lambda \phi$, substituting the gauge parameter by the infinitesimal dressing, $\lambda \rarrow \upsilon$.
The perturbatively dressed fields are $\K$-invariant at first order:
\begin{equation}
\begin{aligned}
\delta_\lambda (\phi^\upsilon)
&= \delta_\lambda \phi + \b\delta_{(\delta_\lambda \upsilon)} \phi 
= \delta_\lambda \phi + \b\delta_{-\lambda} \phi \\
& = \delta_\lambda \phi - \delta_\lambda \phi \equiv 0,
\end{aligned}
\end{equation}
neglecting higher-order terms  in  $\lambda$ and $\upsilon$.
A perturbatively dressed connection is 
$A^\upsilon := A + \b\delta_\upsilon A = A+ D^A\upsilon$.

Considering a  quasi-invariant Lagrangian, i.e. such that
$\delta_\lambda L(\phi)= dB( \phi; \lambda)$, its perturbative dressing is
\begin{align}
\label{pert-dressed-Lagrangian}
  L( \phi^\upsilon):= L( \phi) + dB( \phi; \upsilon). 
\end{align}
The field equations $\bs E( \phi^\upsilon)=0$ are thus  $\K$-invariant at first order (in both $\lambda$ and $\upsilon$).

We stress that, despite a superficial formal similarity, dressed fields $\phi^u$ are \emph{not} gauge transformed fields $\phi^g$. 
This is clear from the definition of a dressing field, which implies $u \notin \H$.  
The dressing operation is not a mapping from field space $\Phi$ to itself, but a mapping from field space  to another mathematical space: the space of dressed fields $\Phi^u$.
A fortiori, a dressing operation \emph{is not} a gauge fixing. 
See \cite{Berghofer:2024plf, Francois:2024laf,Francois:2024rdm} for more on this point.

\subsection{Ususy matter ansatz via dressing}
\label{Ususy matter ansatz via dressing}

We now establish that 
the spin-$1/2$ field appearing in the matter ansatz \eqref{matterans} is actually a partially invariant variable obtained via the DFM.

To this aim, we start with the most  general $\mathfrak{osp}(2|2)$ superconnection $\mathbb{A}_{\mathfrak{osp}}
=\frac{1}{2}\omega^{ab} \mathbb{J}_{ab}+  \bar{\mathbb{Q}} \psi - \bar \psi \mathbb{Q}+A\, \mathbb{T}$, 
with curvature $\mathbb{F}_{\mathfrak{osp}}=d\mathbb{A}_{\mathfrak{osp}} + \tfrac{1}{2}[\mathbb{A}_{\mathfrak{osp}}, \mathbb{A}_{\mathfrak{osp}}]
=d\mathbb{A}_{\mathfrak{osp}}
+\mathbb{A}_{\mathfrak{osp}}^2$.
In our compact notation it reads
\begin{align}
\label{connectionosp}
    \mathbb{A}_{\mathfrak{osp}} 
= \begin{pmatrix}
  \omega &\  \tfrac{1}{\sqrt{\ell}}\,\psi \\[2mm]
   - \tfrac{1}{\sqrt{\ell}}\,\bar \psi  & \  A \otimes J
\end{pmatrix},
\end{align}
and it gauge-transforms as
\begin{align}
\label{infgaugetrosp}
&\delta_\lambda \mathbb{A}_{\mathfrak{osp}} 
= \begin{pmatrix}
    \delta_\lambda \omega  & 
 \ \tfrac{1}{\sqrt \ell} \delta_\lambda \psi  \\[2mm]
 -\tfrac{1}{\sqrt \ell} \delta_\lambda   \bar \psi &\    \delta_\lambda A \otimes J
    \end{pmatrix} = D^{\mathbb A} \lambda 
\\[2mm]
&={\scriptsize{\begin{pmatrix}
 d\beta + [\omega,\beta] - \tfrac{1}{\ell}(\psi\,\bar\varepsilon - \varepsilon \,\b\psi) 
 & \tfrac{1}{\sqrt \ell}\left[
 \nabla \varepsilon
 - (\beta-\alpha \otimes J) \psi \, \right] \\[2mm]
 *
  & \left( d\alpha+ \tfrac{1}{2\ell} \mathrm{Tr} (\psi\, \bar \varepsilon - \varepsilon \, \b\psi ) \right) \otimes J 
\end{pmatrix} .}} \notag 
\end{align}
We first notice that the components of the general 1-form field $\psi=\psi_\mu dx^\mu$ have the following (reducible) ``gamma-trace" decomposition:
\begin{align}
\label{gammatracedec}
\psi_\mu (\rho, \zeta) = \rho_\mu + i \gamma_\mu \zeta ,
\end{align}
where $\rho_\mu$, satisfying $\gamma^\mu \rho_\mu =0$, contains both a ``longitudinal" (divergence-free) mode and a ``transverse" mode, and $\zeta:=- \frac{i}{3} \gamma^\mu \psi_\mu$ is a spin-$1/2$ field.  
This decomposition generalises
to $d$ spacetime dimensions as  $\zeta:=- \frac{i}{d} \gamma^\mu \psi_\mu$. 
By \eqref{infgaugetrosp}, the  susy transformation of $\psi$ is $\delta_\varepsilon \psi = \nabla \varepsilon$, in components $\delta_\varepsilon \psi_\mu = \nabla_\mu \varepsilon$, 
and given \eqref{gammatracedec} it splits as
\begin{equation}
\begin{aligned}
\label{susytrzetarho}
    \delta_\varepsilon \zeta & = - \frac{i}{3} \slashed{\nabla} \varepsilon , \\
    \delta_\varepsilon \rho_\mu & = \frac{i}{3} \gamma_\mu \slashed{\nabla} \varepsilon + \nabla_\mu \varepsilon
    =: -b_\mu (\varepsilon), 
\end{aligned}
\end{equation}
where we define the $\left(\mathfrak{spin}(1,\!2)\!\oplus\! \mathfrak{u}(1)\right)$-covariant operator $b_\mu:=- \frac{i}{3} \gamma_\mu \slashed{\nabla} - \nabla_\mu$, whose formal left inverse is $[b\-]^\mu$.
We now aim to build the susy-invariant dressed field $\psi_\mu^\upsilon := \psi_\mu+ \nabla_\mu \upsilon$,
where $\upsilon=\upsilon[\psi]$ is a spinorial susy dressing field. 
The gamma-trace decomposition of  $\psi_\mu^\upsilon $ is 
\begin{equation}
\begin{aligned}
\label{gamma-tr-dressed-psi}
\psi_\mu^\upsilon=&\, \rho_\mu^\upsilon + i \gamma_\mu \zeta^\upsilon\\
:=&\, \left( \rho_\mu - b_\mu(\upsilon) \right) +i\gamma_\mu \left(\zeta - \tfrac{1}{3}\slashed \nabla \upsilon \right).
\end{aligned}
\end{equation}
The dressing field is found by solving explicitly the constraint $\rho_\mu^\upsilon=0$, so that, indeed,
\begin{align}
\label{susy-dressing}
  &\upsilon[\psi]:= [b\-]^\mu(\rho_\mu) \quad \text{and } \\[1mm]
  & 
  \delta_\varepsilon \upsilon := \upsilon[\delta_\varepsilon \psi]
  \approx [b\-]^\mu (\delta_\varepsilon\rho_\mu)
  = [b\-]^\mu(-b_\mu (\varepsilon))
  = -\varepsilon , \notag
\end{align}
as expected by definition \eqref{dressingfieldtr} of an infinitesimal dressing field. 
We therefore end up, as intended, with the susy-invariant dressed field
\begin{align}
 \label{dressed-psi}
 \psi_\mu^\upsilon = i\gamma_\mu \zeta^\upsilon =: i\gamma_\mu \chi,
\end{align}
where we defined the susy-invariant dressed spinor 
$\chi:= \zeta^\upsilon := \zeta - \frac{1}{3} \slashed{\nabla} \upsilon$.
We thus obtain the matter ansatz \eqref{matterans} of the AVZ model via the DFM \eqref{dressed-psi}, producing its spin-$1/2$ field $\chi$ as a susy-invariant dressed variable.

Let us now make a few important observations. 
First, since $\upsilon=\upsilon[\psi]$ in \eqref{susy-dressing} is non-local -- involving the formal inverse of a differential operator  -- so is the dressed field \eqref{dressed-psi}. 
In  the terminology of \cite{Francois:2017akk,Francois:2024rdm}, we might thus say that susy is a ``substantive" symmetry, as it is reduced via dressing at the  cost of locality. 

Second, the condition 
$\rho_\mu^\upsilon=0$ is susy-invariant, and is \emph{not} a gauge fixing of $\psi_\mu$. 
As per the usual caveat expressed in the DFM, $\psi^\upsilon$ is not a gauge-fixed version of $\psi$: being susy-invariant, the former does not even belong to the same mathematical space as the latter.

Finally, and relatedly, we stress that 
for field-dependent dressing fields $\upsilon=\upsilon[\phi]$, dressed fields $\phi^\upsilon$ have a natural interpretation as \emph{relational variables}: they encode the gauge-invariant \emph{relations} among physical d.o.f.. 
Said otherwise, the DFM is a systematic tool to implement the idea  that physical d.o.f. invariantly coordinatise each other -- which encompasses e.g. scalar coordinatisation in General Relativity, à la Kretschmann-Komar \cite{Komar1958} or Brown-Kucha\v{r} \cite{Brown-Kuchar1995}. 
As observed in the introduction, relationality thus understood is the conceptual core of general-relativistic gauge field theory, including supersymmetric field theory, see \cite{Francois:2024vlr, Francois:2024laf, Francois:2024rdm, FrancoisAndre:2023jmj}. 
The~dressed spinor $\chi=\zeta^\upsilon$ is such a relational variable, encoding the susy-invariant physical  relations among the d.o.f. embedded in the bare field $\psi$. 

To fully exploit the power of the DFM, we can now dress the full $\mathfrak{osp}(2|2)$ superconnection
\eqref{connectionosp}. 
Writing the dressing field in matrix form as 
\begin{align}
 \label{matrix-dressing}
 \bs\upsilon = 
 \begin{pmatrix}
    0 & \ \tfrac{1}{\sqrt \ell} \upsilon \\[1mm]
    - \tfrac{1}{\sqrt \ell} \b\upsilon\  & 0
 \end{pmatrix},
\end{align}
by \eqref{pert-dressed-fields}, the susy-invariant dressing of $\mathbb{A}_{\mathfrak{osp}}$ is 
\begin{equation}
\begin{aligned}
\label{dressed-osp-connection}
\mathbb{A}_{\mathfrak{osp}}^\upsilon :=&\, \mathbb{A}_{\mathfrak{osp}} +\b\delta_{\bs \upsilon}\mathbb{A}_{\mathfrak{osp}} \\
=&\, \mathbb{A}_{\mathfrak{osp}} + D^{\mathbb{A}_{\mathfrak{o\!s\!p}}} \bs \upsilon  \\[1mm]
=&\,
\begin{pmatrix}
  \omega^\upsilon &\  \tfrac{1}{\sqrt{\ell}}\,\psi^\upsilon \\[2mm]
   - \tfrac{1}{\sqrt{\ell}}\,\bar \psi^\upsilon  & \  A^\upsilon \otimes J
\end{pmatrix}  \\
=&\, 
\begin{pmatrix}
  \omega^\upsilon &\  \tfrac{i}{\sqrt{\ell}}\,\gamma \chi \\[2mm]
    \tfrac{i}{\sqrt{\ell}}\,\bar \chi \gamma  & \  A^\upsilon \otimes J
\end{pmatrix}  \\
{\scriptsize{=}}&{\scriptsize{\,\,\begin{pmatrix}
 \omega - \tfrac{1}{\ell}(\psi\,\bar\upsilon - \upsilon \,\b\psi) 
 & \tfrac{1}{\sqrt \ell}\left[
 \psi+ \nabla \upsilon \right] \\[2mm]
 *
  & \left( A + \tfrac{1}{2\ell} \mathrm{Tr} (\psi\, \bar \upsilon - \upsilon \, \b\psi ) \right) \otimes J 
\end{pmatrix} ,}} 
\end{aligned}
\end{equation}
which is none other than a susy-invariant version of the AVZ connection \eqref{connection}. 
The associated invariant dressed curvature is 
$\mathbb{F}_{\mathfrak{osp}}^{\,\upsilon} := \mathbb{F}_{\mathfrak{osp}} +\b\delta_\upsilon \mathbb{F}_{\mathfrak{osp}}
= \mathbb{F}_{\mathfrak{osp}} + [\mathbb{F}_{\mathfrak{osp}},  \bs\upsilon]$. 
Observe how the invariant $\mathfrak{spin}(1,2)$-connection $\omega^\upsilon$ and 
$\mathfrak{u}(1)$-connection $A^\upsilon$ are dressed with d.o.f. from $\psi$ via $\upsilon$, and therefore encode the relational d.o.f. between the bare fields $\omega$, $A$, and $\psi$.
 
The dressing of the CS Lagrangian  $L (\mathbb A_{\mathfrak{osp}})$ for $\mathbb A_{\mathfrak{osp}}$ is, by \eqref{pert-dressed-Lagrangian},
\begin{equation}
\begin{aligned}
\label{dressed-CS-lag}
    L (\mathbb A_{\mathfrak{osp}}^\upsilon)
    &=  L (\mathbb A_{\mathfrak{osp}}) + dB( \mathbb A_{\mathfrak{osp}} ; \upsilon) \\
    &= 
     \mathrm{sTr} \left(\mathbb{A}^\upsilon_{\mathfrak{osp}} \mathbb{F}^{\,\upsilon}_{\mathfrak{osp}} - \frac{1}{3} (\mathbb{A}^\upsilon_{\mathfrak{osp}})^3 \right). 
\end{aligned}
\end{equation}
It is the (u)susy-invariant version of the AVZ Lagrangian \eqref{CSLagr}. 
The (u)susy-invariant dressed field equations are then $\mathbb{F}^{\,\upsilon}_{\mathfrak{osp}}=0$, which unfold as dressed versions of \eqref{unfolded-field-eq}. 

We stress again that, since a dressing is not a gauge fixing, the dressed Lagrangian \eqref{dressed-CS-lag} is not a gauge-fixed version of $L (\mathbb A_{\mathfrak{osp}})$. 
Rather, $L (\mathbb A^\upsilon_{\mathfrak{osp}})$ is the relational rewriting of the theory; its field equations $\mathbb{F}^{\,\upsilon}_{\mathfrak{osp}}=0$ are the relational version of the bare ones $\mathbb{F}_{\mathfrak{osp}}=0$, and they differ only by a boundary term -- a general feature proven in full generality in \cite{Francois:2024rdm}.

\section{Residual gauge symmetries}\label{Residual gauge symmetries}

We analyse the  gauge transformations remaining after dressing. 
First, we focus on the natural residual Lorentz and $\U(1)$  symmetries of the model. 
Then we comment on the so-called Nieh-Yan-Weyl symmetry \cite{Nieh:1981xk,Chandia:1997hu,Kreimer:1999yp,Hughes:2012vg,Parrikar:2014usa} that has been associated to the matter ansatz \eqref{matterans}/\eqref{dressed-psi}. 

\subsection{Residual Lorentz and $\U(1)$ transformations}
\label{Residual Lorentz and U(1) transformations}

Since the  odd, susy, part of the gauge superalgebra Lie$\mathcal{OS}p(2|2)$ has been reduced via dressing, one expects that the dressed fields $\mathbb{A}^{\,\upsilon}_{\mathfrak{osp}}$, and $\mathbb{F}^{\,\upsilon}_{\mathfrak{osp}}$, exhibit residual transformations under the remaining even gauge subalgebra 
Lie$\mathcal{S}pin(1,2)\,\oplus$ Lie$\mathcal U(1)$, with parameter 
\begin{align}
 \label{residual-sym}  
 \theta =\begin{pmatrix}
     \beta\ & 0 \\[1mm]
     0 \ & \alpha \otimes J
 \end{pmatrix}.
\end{align}
A basic result of the DFM is that if one proves  
\begin{align}
\label{residual-trsf-dressing}
 \delta_\theta \bs\upsilon &= [\bs\upsilon, \theta], \\[1mm]
 \begin{pmatrix}
     0\ & \tfrac{1}{\sqrt \ell} \delta_\theta \upsilon \\[1mm]
     -\tfrac{1}{\sqrt \ell} \delta_\theta \b\upsilon \ & 0 
 \end{pmatrix}
 &=
 \begin{pmatrix}
     0\ & -\tfrac{1}{\sqrt \ell} (\beta - \alpha\otimes J) \upsilon \\[1mm]
     * \ & 0 
 \end{pmatrix} ,  \notag
\end{align}
then it follows that  
\begin{equation}
 \label{residual-trsf-dressed-fields}
\begin{aligned}
 \delta_\theta \mathbb{A}^{\,\upsilon}_{\mathfrak{osp}} &= D^{\mathbb{A}^{\upsilon}_{\mathfrak{o\!s\!p}}}\theta
 = d\theta + [\mathbb{A}^{\,\upsilon}_{\mathfrak{osp}}, \theta]
 , \\[1mm]
 \text{ and } \ 
 \delta_\theta \mathbb{F}^{\,\upsilon}_{\mathfrak{osp}} &= [\mathbb{F}^{\,\upsilon}_{\mathfrak{osp}}, \theta ].
\end{aligned}
\end{equation}
That is, the susy-invariant fields have standard residual Lorentz and $\U(1)$ gauge transformations. Which implies that the field equations have the required residual covariance.
To secure the result, one needs only to prove \eqref{residual-trsf-dressing}:
The dressing has been found by solving $\rho_\mu^\upsilon=0$, i.e. 
$\rho_\mu = b_\mu(\upsilon)$. And by  \eqref{infgaugetrosp}-\eqref{gammatracedec} we have that $\delta_\theta \rho_\mu =-(\beta - \alpha \otimes J) \rho_\mu$. Now, since $b_\mu$ is a $\left(\mathfrak{spin}(1,\!2)\!\oplus\! \mathfrak{u}(1)\right)$-covariant operator, the latter holds if $\upsilon$ has the same covariance as $\rho_\mu$; so $\delta_\theta \upsilon =-(\beta -\alpha\otimes J)\upsilon$, as required in \eqref{residual-trsf-dressing}. So, indeed, \eqref{residual-trsf-dressed-fields} holds.

\subsection{NYW invariance as an artificial symmetry}
\label{NYW invariance as an artificial symmetry}

The  matter ansatz \eqref{matterans} is invariant under local scale transformations associated with the so-called NYW symmetry \cite{Nieh:1981xk,Chandia:1997hu,Kreimer:1999yp,Hughes:2012vg,Parrikar:2014usa}. In $d$ spacetime dimensions, in terms of the scaling parameter $\Xi(x)$,  the NYW transformation are:
\begin{equation}
\label{nywn}
\begin{aligned}
    & e^a_\mu \mapsto e^{\Xi(x)} e^a_\mu , \\   
    & \chi \mapsto e^{-\tfrac{d-1}{2} \Xi(x)} \chi . 
\end{aligned}
\end{equation}
In the AVZ model of ususy, the NYW symmetry is identified as the Weyl symmetry associated with a conformal rescaling of the metric on the base space.
In $d=3$,
\begin{align}
\label{nyw3}
     \chi \mapsto e^{- \Xi(x)} \chi , 
\end{align}
and the NYW transformations \eqref{nywn}-\eqref{nyw3} clearly leave $\psi$ in \eqref{matterans} invariant, or $\psi^\upsilon$ in  \eqref{dressed-psi}.
So, since $\omega$ and $A$ are NYW singlets, $\mathbb A$ and $\mathbb A_{\mathfrak{osp}}^\upsilon$ are NYW invariant, and therefore so is the model. 

Yet, the transformations \eqref{nywn}-\eqref{nyw3} are external to the $\rm{OSp}(2|2)$ supergeometry underlying the model, they are imposed on it by  fiat and therefore carry little relevant information about its kinematics. 
Furthermore, since it is a symmetry that can be ``reduced" without losing the locality of the theory -- via dressing, considering ${e^a}_\mu$ as a dressing field for $\chi$ --  in the terminology of \cite{Francois:2017akk, Francois:2024rdm} it is an ``artificial" symmetry, or ``fake" symmetry in the terms of \cite{Jackiw-Pi2015}.
So, if the NYW transformations \eqref{nywn}-\eqref{nyw3} can  be thought of as a residual symmetry of the dressed theory $L(\mathbb{A}^\upsilon_\mathfrak{osp})$, it is an artificial residual symmetry.

We observe that it is analogous to the so-called ``Stueckelberg symmetry" arising from the \emph{Chen et al ansatz} \cite{Chen:2008ag,Chen:2009mr}, related to the issue of finding a gauge-invariant decomposition of the angular momentum operator of nucleons \cite{Leader:2013jra}. 
An issue clarified by the DFM~\cite{FLM2015_I}.

\section{Conclusions}\label{Conclusions}

In this work we had two goals, 1. and 2. 
\mbox{Achieving}~1. entailed to show first how the \emph{matter ansatz}, on which rests the ususy approach to a unified description of gauge and matter fields within a single superconnection, 
emerges as a result of the application of the Dressing Field Method -- thereby correcting previous literature \mbox{interpreting} the ansatz as a mere gauge fixing. 

For concreteness, we illustrated our claim by thoroughly working out the historically first and simplest ususy model, the AVZ model based on the supergroup $\rm{OSp}(2|2)$. 
To do so efficiently, we have first clarified the structure of the gauge supergroup and superalgebra arising from the bundle supergeometry underlying the model, and then exploited a new compact matrix notation, greatly streamlining the presentation of the model and clarifying the conceptual picture.
We stress that such a matrix notation allows to effectively deploy the full power of the DFM -- see also in that regard its applications in conformal and twistor geometry \cite{FLM2016_I, Attard-Francois2016_I, Attard-Francois2016_II, Francois2019} which elucidated the geometric origin of the field space underlying the algebraic classification of Weyl anomalies in conformal field theory \cite{PhysRevLett.98.261302}. 
We then applied the DFM to the most general $\mathcal{OS}p(2|2)$ gauge theory with connection $\mathbb A_{\mathfrak{osp}}$: the Grassmann-odd component of the susy-invariant dressed connection $\mathbb A_{\mathfrak{osp}}^\upsilon$  is precisely of the form \eqref{matterans}/\eqref{dressed-psi} of the matter ansatz. 
Through the DFM framework, $\mathbb A_{\mathfrak{osp}}^\upsilon$  is furthermore easily shown to behave as a standard gauge potential under the residual bosonic gauge subgroup $\mathcal{S}pin(1,2)\times\mathcal U(1)$ of $\mathcal{OS}p(2|2)$, giving the correct gauge transformations for the Lorentz and $\rm{U}(1)$ connections and the spin-$1/2$ field.

Elucidating the technical and conceptual status of the ansatz opens the possibility to extend in a principled way, and in conjunction with the DFM, the ususy proposal beyond the above special circumstances of its inception. 
It can thereby become a more general approach to a unification of gauge and matter fields in dimension $d\geq3$ within the framework of supersymmetric field theory, distinct from standard and more speculative  high energy physics applications of the latter. 
This in a way goes back to the root of Berezin's original motivation for the introduction of supergeometry in fundamental physics.  
We  indicate how  to proceed in Appendix \ref{Ususy via DFM in any dimension} hereafter, cementing our goal 1., and stressing the root reasons for the failure of previous attempts:  i.e.  ``wrong" choices of superalgebras  resulting from a misappreciation of the core idea of the ususy program and of a lack of awareness of the conceptual key afforded by Cartan (super-)geometry \cite{Francois:2024rfh}.

As noted already, interesting parallels are then to be drawn with non-commutative geometric approaches to the unification of the gauge and Higgs fields.  
A fusion of both endeavours, realized e.g. in the framework  of super-Lie algebroids, would lead to geometrically unified Einstein-Yang-Mills-Dirac-Higgs models, where gauge, matter and Higgs fields are  parts of a generalised super-connections.
Evaluating the feasibility and fruitfulness of such a program is left for subsequent works.

Achieving our second goal 2. meant to showcase 
the DFM as a key ingredient in this now opening ususy program, giving a clean template for how it applies 
as a way to introduce it
to the broader community of researchers  working within the framework of general-relativistic~gauge~field~theory. 
Technically, the DFM is a new systematic approach to   rewrite a theory in a manifestly gauge-invariant way, thereby considerably simplifying the analysis of its physical content.
Conceptually, it is perfectly resonant with the (often misappreciated) core paradigmatic insight of general-relativistic gauge field theory: relationality \cite{Francois:2024vlr}. 
Indeed, in the DFM gauge-invariance is achieved via production of  physical \emph{dressed} d.o.f. representing relations among \emph{bare} d.o.f.  \cite{FrancoisAndre:2023jmj,Francois:2024rdm}. 
In other words, physical d.o.f. arise from the  mutual and invariant coordinatisation of gauge-variant d.o.f. by each other. 
The insight naturally exports to supersymmetric field theory, as shown here, and showcased by the susy-invariant dressed field $\mathbb A_{\mathfrak{osp}}^\upsilon$ above. 
See also  \cite{Francois:2024laf}, where  the Rarita-Schwinger and gravitino fields are also shown to be dressed relational variables, and not gauge-fixed as is often claimed. 
The clear \mbox{mathematical} \mbox{difference} \mbox{between} dressing and gauge fixing should be kept in sight, as many instances of popular ``gauges", e.g. the Coulomb and Lorentz gauges in electromagnetism, or the unitary gauge in the electroweak model, were shown to actually be cases of dressings \cite{Berghofer:2024plf,Francois:2024rdm} -- so is the case for, e.g.,  the axial gauge, the harmonic gauge, or the de Donder gauge.
Furthermore, it bears stressing that what is  confronted to observations and experiments is never a ``gauge-fixed" theory, but its relational dressed version \cite{Francois:2024rdm}.  

The DFM has therefore broad applications in general-relativistic gauge field theory, and we expect that many fruitful and clarifying instances are yet to be found across the field, from high energy to condensed matter physics.

\begin{acknowledgments}
We thank Philipp Berghofer for useful feedback on the manuscript.
J.F. is supported by the Austrian Science Fund (FWF), grant \mbox{[P 36542]}, 
and by the Czech Science Foundation (GAČR), grant GA24-10887S.
L.R. is supported by the research grant GrIFOS, funding scheme PNRR Young Researchers, MSCA Seal of Excellence (SoE), CUP E13C24003600006, ID SOE2024$\_$0000103, of which this paper is part.
She also thanks the Department of Physics and the Department of Mathematics and Statistics at the Faculty of Science of MUNI for the kind hospitality during her stay in which the idea behind this work was conceived. 
\end{acknowledgments}

\appendix

\section{Ususy via DFM in any dimension}
\label{Ususy via DFM in any dimension}

What is done in Sections \ref{The AVZ model of ususy} to \ref{Residual gauge symmetries} generalises to other models in $d>3$ whereby gauge and matter fields can be unified in a (dressed) super-connection, fulfilling the key idea of the ususy proposal as we think is best construed. 
To see why, and how, one must think from first principles. 

First, one must clearly distinguish the kinematics -- where the DFM is to be applied -- from the dynamics, keeping in mind that a  kinematics is given by bundle supergeometry.
Then, one must understand that the ususy program is at core a requirement of  \emph{specific \mbox{kinematics}}: one given by a superbundle with bosonic base $M$ and superspace fibers, implying a gauge super-group $\H$, i.e. \emph{internal} supersymmetry. 
The complete set of fields are thus the canonical soldering form $e^a$ of the base and a (non-canonical) Ehresmann super-connection $\mathbb{A}$; these are subject to the full local symmetry group  $\Diff(M)\ltimes \H$. 

The roots of the limitations of  existing literature \mbox{attempting} to extend ususy --  e.g. to $4d$ models as in 
\cite{Alvarez:2013tga, Alvarez:2020qmy, Alvarez:2021zhh, Alvarez:2023auf} -- is that A) it did not pay close enough attention  to the conceptual structure just laid,~and/or B)~it~overlooked the \emph{Cartan geometric} structure underlying a compelling gauge theory of gravity \cite{Francois:2024rfh, Sharpe, Cap-Slovak09}. 
This~resulted in choices of superalgebras $\mathfrak g$ for the kinematics where, like the super-Poincaré algebra,  the brackets of susy generators $\mathbb{Q}$ have a component along the translations (or transvections) generator $\mathbb{P}$: 
$\mathfrak{osp}(4|2)$ in \cite{Alvarez:2013tga}, 
$\mathfrak{su}(2,2|2)$ and $\mathfrak{su}(2,2|N)$ in \cite{Alvarez:2020qmy, Alvarez:2021zhh, Alvarez:2023auf}.
This leaves with either of two options: Either,
I) to treat the $\mathfrak g$-valued superconnection as an Ehresmann connection, acted upon by a gauge group $\mathcal G$. A move exactly analogous to \emph{Poincaré gauge gravity} or \emph{Metric-Affine Gravity} (MAG), and suffering all the same technical drawbacks and conceptual issues; chief among which the presence of unwanted ``internal gauge translations" 
and the presence of a ``translational gauge potential" somehow to be identified with (or made to induce) the 
vierbein
on $M$ \cite{footnote1}.
Or, 
II) more sensibly, to consider the Cartan geometric picture natural for these algebras: 
i.e. if finding $\mathfrak{h}\subset \mathfrak{g}$ s.t.  $(\mathfrak{g}, \mathfrak{h})$ is a  super-Klein pair on which a Cartan supergeometry is modeled, $H$ being the fibration and the vielbein being $\mathfrak{g}/\mathfrak{h}$-valued. 
In the cases mentioned, this unfortunately gives a superbundle over a \emph{supermanifold}~$\b M$ rather than over a manifold $M$, so that the spin-$3/2$ field is part of a supervielbein and  supersymmetry is the odd part of  $\Diff(\b M)$ rather than an internal gauge symmetry. 
We are there in the land of ``standard" supergravity, with the attending problems necessitating to restrict fields of $\b M$ as fields on $M \subset \b M$. 
A context in which the DFM applies \cite{Francois:2024laf}, but which cannot lead to a compelling ususy unification of matter and gauge fields. 

The problem was thus not that ususy is tied to  $d=3$ and encounters problems in $d>3$;  spacetime dimension is only tangential to the core issues just articulated.  
\medskip

To obtain sensible ususy models, one needs only  to choose a superalgebra according to the kinematical desiderata we have highlighted, while being  mindful of a possible Cartan supergeometric structure. 
One may e.g. pick from the classification by Kac \cite{Kac1977}. 
Applying the DFM, one may then obtain  a susy-invariant dressed super-connection featuring a ``matter ansatz", i.e.  gauge and  matter fields unified. 
\medskip

For example, Lorentz superalgebras fit our template.
To illustrate, let us choose the simplest possible example in $d=4$: the semi-direct extension $\LieH=\mathfrak{sl}(2, \CC) \oplus \CC^2$ of the Lorentz algebra $\mathfrak{spin}(1,3)\simeq\mathfrak{sl}(2, \CC)$ by an (additive) Abelian superalgebra. 
It is all but analogous to the Poincaré algebra, but for translation generators in the fundamental representation of the Lorentz algebra being replaced by susy generators in its spin representation.

We have thus a superbundle with structure supergroup $H=SL(2, \CC) \ltimes \CC^2$ over an even manifold $M$, with gauge supergroup $\H:= \{ g, g':  M \rarrow H\,|\, {g}^{\prime\, g} =g\- g' g \}$ and  gauge algebra
Lie$\H:= \{ \lambda, \lambda': M \rarrow~\LieH\,|\, \delta_\lambda \lambda' = [\lambda', \lambda] \}$. 
In matrix form, we have
\begin{align}
\label{gaugeparam-trsf-ex1}
 \delta_\lambda\lambda' 
&= \begin{pmatrix}
    \delta_\lambda \beta'  & 
 \  \delta_\lambda \varepsilon' \\[2mm]
0  &\    0
    \end{pmatrix} 
:= \begin{pmatrix}
 [\beta', \beta] &  \beta ' \varepsilon  - \beta \varepsilon'\\[2mm]
 0  & 0
\end{pmatrix},
\end{align}
with $\beta \in \mathfrak{sl}(2, \CC)$ and $\varepsilon \in \CC^2$ a spinorial susy parameter. 
A gauge group element is $g=\begin{pmatrix} s & \varsigma\\ 0 & 1 \end{pmatrix}$, with $s\in \SL(2, \CC)$ and $\varsigma:M \rarrow  \CC^2$ a susy spinor, and the semi-direct structure of $\H$ is reproduced by matrix multiplication. 

An Ehresmann superconnection and its curvature are
\begin{align}
\label{connection-ex1}
    \mathbb{A} 
= \begin{pmatrix}
  \omega &\   \psi \\[2mm]
   0  & \  0
\end{pmatrix}, \quad
\mathbb{F} 
= \begin{pmatrix}
  \Omega &\   \varrho \\[2mm]
   0  & \  0
\end{pmatrix}
=\begin{pmatrix}
  d\omega+\omega^2 &\   d\psi+ \omega \psi \\[2mm]
   0  & \  0
\end{pmatrix}. \notag
\end{align}
The curvature being $\mathbb F=d \mathbb{A}  + \tfrac{1}{2}[\mathbb{A} , \mathbb{A} ] = d \mathbb{A} + \mathbb{A}^2$, it satisfies Bianchi identity $D^\mathbb{A}\mathbb{F}= d\mathbb{A} +[\mathbb{A}, \mathbb{F}]=0$.
Their  Lie$\H$-transformations are
\begin{equation}
\begin{aligned}
    \delta_\lambda \mathbb A &= D^\mathbb{A}\lambda
    =\begin{pmatrix}
      \nabla \beta  & \ -\beta \psi+ \nabla \varepsilon  \\ 0 & 0 
  \end{pmatrix} , \\[2mm]
     \delta_\lambda \mathbb F &= [\mathbb{F}, \lambda]
     =\begin{pmatrix}
      [\Omega, \beta]  & \ -\beta \varrho+ \Omega \varepsilon  \\ 0 & 0 
  \end{pmatrix},
\end{aligned}
\end{equation}
with $\nabla \beta = d\beta + [\omega ,\beta]$ and $\nabla \varepsilon = d\varepsilon + \omega \varepsilon$.
This achieves to define the kinematics that is soon to provide that of our simplest $4d$ ususy model.

Then, one uses the very same ``gamma-trace" decomposition \eqref{gammatracedec} of the 1-form $\psi$ as in the AVZ model, from which we extract the same dressing field $\upsilon=\upsilon[\psi]$ \eqref{susy-dressing}, which may be cast in matrix form as 
\begin{align}
 \label{matrix-dressing-ex1}
 \bs\upsilon = 
 \begin{pmatrix}
    0 & \  \upsilon \\[1mm]
    0 & 0
 \end{pmatrix}.
\end{align}
It allows to build the susy-invariant dressed fields
\begin{equation}
\label{1}
\begin{aligned}
\mathbb{A}_{\mathfrak{}}^\upsilon :=&\, \mathbb{A}_{\mathfrak{}} +\b\delta_{\bs \upsilon}\mathbb{A}_{\mathfrak{}} 
=\, \mathbb{A}_{\mathfrak{}} + D^{\mathbb{A}_{\mathfrak{}}} \bs \upsilon , \\[1mm]
\begin{pmatrix}
     \omega^\upsilon &\ \psi^\upsilon \\[1mm]  0 & 0
\end{pmatrix}=&\,
\begin{pmatrix}
 \ \omega &\  \psi + \nabla \upsilon \\[1mm]
   0  &\ 0
\end{pmatrix} 
=
\begin{pmatrix}
 \ \omega &\  i \gamma \chi \\[1mm]
   0  &\ 0
\end{pmatrix},
\\[3mm]
\mathbb{F}_{\mathfrak{}}^{\,\upsilon} :=&\, \mathbb{F}_{\mathfrak{}} +\b\delta_{\bs \upsilon}\mathbb{F}_{\mathfrak{}} 
= \mathbb{F}_{\mathfrak{}} + [\mathbb{F}_{\mathfrak{}}, \bs \upsilon ] , \\[1mm]
\begin{pmatrix}
     \Omega^\upsilon &\ \varrho^\upsilon \\[1mm]  0 & 0
\end{pmatrix}=&\,
\begin{pmatrix}
 \ \Omega &\  \varrho + \Omega \upsilon \\[1mm]
   0  &\ 0
\end{pmatrix}.
\end{aligned}
\end{equation}
There, $\gamma = \gamma_\mu dx^{\,\mu}=\gamma_a {e^a}_\mu dx^{\,\mu} =\gamma_a e^a$ is the gamma-matrix 1-form, $e^a$ is the vierbein 1-form of $M$,
and we have the  $\CC^2$-spinor field $\chi=\zeta^\upsilon = \zeta - \tfrac{1}{4} \nabla \upsilon$, with $\zeta=-\tfrac{i}{4}\gamma^\mu \psi_\mu$ analogous to \eqref{gamma-tr-dressed-psi}. 
Hence, a ``matter ansatz" is realized  via dressing, so that gauge and matter fields feature on equal footing in the susy-invariant superconnection $\mathbb A^\upsilon$, as expected from a ususy model. 

Because the dressing field $\upsilon$ is non-local, so are the dressed fields \eqref{1}. This indicates that the susy gauge symmetry is (here again) substantive: its elimination cost the field locality. This trade-off ``invariance \emph{vs} locality" is a typical signature of substantive gauge symmetries \cite{Francois:2017akk, Francois:2024rdm}.
We may also stress that since $\upsilon=\upsilon[\psi]$, the susy-invariant fields \eqref{1} are \emph{relational} variables. 

Regarding the residual Lie$\SL(2, \CC)$-transformations, as stated in section \ref{Residual Lorentz and U(1) transformations},
if one can show that 
\begin{equation}
\label{A}
\begin{aligned}
  \delta_\beta \bs\upsilon &= [\bs\upsilon, \bs\beta], \\[1mm]
  \begin{pmatrix}
      0 & \delta_\beta \upsilon \\ 0 & 0 
  \end{pmatrix}
  &=
  \begin{pmatrix}
      0 & -\beta \upsilon \\ 0 & 0
  \end{pmatrix},
\end{aligned}
\end{equation}
where $\bs \beta = \begin{pmatrix}
    \beta & 0 \\ 0 & 0
\end{pmatrix}$,
then it follows that  
\begin{align}
\label{B}
 \delta_\beta \mathbb{A}^{\upsilon} &= D^{\mathbb{A}^{\upsilon}}\bs\beta
 = d\bs\beta + [\mathbb{A}^{\upsilon}, \bs\beta]=\begin{pmatrix}
     d\beta + [\omega, \beta] &\ -i\,\beta \gamma \chi \\ 0 & 0
 \end{pmatrix}, \notag\\[1mm] 
 \delta_\theta \mathbb{F}^{\,\upsilon} &= [\mathbb{F}^{\,\upsilon}, \bs\beta ].
\end{align}
 That is, the susy-invariant fields \eqref{1} have standard residual Lorentz gauge transformations.
The Lorentz covariance of the dressing field \eqref{A} is proven by the same argument provided in Section \ref{Residual Lorentz and U(1) transformations} below \eqref{residual-trsf-dressed-fields}, whereby the result \eqref{B} hold.

As in section \ref{NYW invariance as an artificial symmetry}, one may insert the identity into the dressed field $\psi^\upsilon=i\gamma\chi$ in the form of the split \eqref{nywn} and call it
a ``NYW symmetry". 
All basic fields \eqref{1} being singlet under this operation, which is imposed by fiat on the kinematics, it  holds no  relevant physical information. 
As a matter of fact, for the reason given in  section \ref{NYW invariance as an artificial symmetry}, this NYW symmetry is \emph{artificial}.
As we have stressed, it is analogous to the
so-called ``Stueckelberg symmetry" arising in the \emph{Chen et al ansatz} \cite{Chen:2008ag,Chen:2009mr} introduced to obtain a gauge-invariant decomposition of  nucleons \cite{Leader:2013jra} spin. 
In both cases the DFM clarifies the technical and conceptual picture \cite{FLM2015_I}.
 
\medskip
All the above is kinematical.
Regarding the dynamics, two possibilities may then be considered: 
Lagrangians that are (quasi-)invariant under either the initial full gauge supergroup $\H$, or  the residual gauge symmetries $\H/$susy ($=\SL(2, \CC)$ here). 
The former case is more constraining, but then it is  possible to dress such a $\H$-invariant Lagrangian according to \eqref{pert-dressed-Lagrangian}  -- as we illustrated by \eqref{dressed-CS-lag} in the case of the AVZ model. 
The latter case allows more freedom, as susy-invariance is kinematically guaranteed by the DFM, the basic dressed fields \eqref{1} being susy-singlets. 
The latter option,  dealing directly with the physical relational d.o.f., is arguably more sensible.
\medskip

We remark that this simple  application of the DFM to a $\mathfrak{sl}(2, \CC) \oplus \CC^2$-valued connection is formally extremely close to the similar treatment via the DFM of Poincaré gauge gravity with $\mathfrak{so}(1, 3) \oplus \RR^4$-valued connection, or in MAG with a $\mathfrak{gl}(n) \oplus \RR^n$-valued connection. 
As recounted in footnote \cite{footnote1}, in these later cases an \emph{ad hoc} dressing $\xi$  for the translation gauge subgroup (sometimes called the ``radius vector") is introduced (by fiat) and plays the exact same role as the susy dressing field $\upsilon$ above -- save for the fact that $\xi$ is local and not built from the connection, so that the dressed connection and curvature are not relational variables.
\medskip

In $d=4$, 
orthosymplectic superalgebras  $\mathfrak{osp}$ containing $\mathfrak{spin}(1,3)\simeq\mathfrak{sl}(2, \CC)\simeq \mathfrak{sp}(2,\CC)$ in their even subalgebra provide another possible class of Lorentz superalgebras that  fit our template -- e.g. $\mathfrak{osp}(2, N; \CC)$ as discussed  in \cite{Abe-Nakanishi1988, Abe1989, ABE1990}. 
This is no surprise since, as seen above, $\mathfrak{osp}(2|2)$ provided such a case in $d=3$. 

Another important class of scenarios fitting our template (in $d\geq 3$) are super-Klein pairs $(\mathfrak{g}, \mathfrak{h})$ with $\mathfrak{h}\subset \mathfrak{g}$ s.t. $\mathfrak{g}/\mathfrak{h}$ is an even vector space, which lead to a ``good" Cartan supergeometric picture: that is, a Cartan superbundle over a bosonic $M$ and a $\mathfrak g$-valued Cartan superconnection whose $\mathfrak{g}/\mathfrak{h}$-part is the vielbein $e^a$ of $M$.
The dressed Cartan superconnection would display $e^a$ both in its $\mathfrak{g}/\mathfrak{h}$-part and in the odd sector of its $\mathfrak h$-part featuring the matter ansatz. 
There is thus no need for ``two metric structures", or  a ``second vielbein", as claimed e.g. in \cite{Alvarez:2013tga, Alvarez:2020qmy, Alvarez:2021zhh, Alvarez:2023auf}.


\bibliography{ususydressing}

\end{document}